\documentclass[twocolumn,showpacs,amsmath,amssymb,pra]{revtex4}

\usepackage{graphicx}
\usepackage{dcolumn}
\usepackage{bm}

\begin{document}


\title {
Electronic and transport properties of rectangular graphene
macromolecules and zigzag carbon nanotubes of finite length}

\author {A. V. Nikolaev$^{1,2}$}

\author{A. V. Bibikov$^1$}

\author{A. V. Avdeenkov$^1$}

\author{I. V. Bodrenko$^1$}

\author{E. V. Tkalya$^1$}

\affiliation{$^1$Institute of Nuclear Physics, Moscow State
University, Vorob'evy Gory, 119992, Moscow, Russia \\
$^2$Institute of Physical Chemistry of RAS, Leninskii pr. 31,
117915, Moscow, Russia}

\date{\today}

\begin{abstract}
We study one dimensional (1D) carbon ribbons with the armchair
edges and the zigzag carbon nanotubes and their counterparts with
finite length (0D) in the framework of the H\"{u}ckel model. We
prove that a 1D carbon ribbon is metallic if its width (the number
of carbon rings) is equal to $2+3n$. We show that the dispersion
law (electron band energy) of a 1D metallic ribbon or a 1D
metallic carbon nanotube has a universal {\it sin-}like dependence
at the Fermi energy which is independent of its width. We find
that in case of metallic graphene ribbons of finite length
(rectangular graphene macromolecules) or nanotubes of finite
length the discrete energy spectrum in the vicinity of
$\varepsilon=0$ (Fermi energy) can be obtained exactly by
selecting levels from the same dispersion law. In case of a
semiconducting graphene macromolecule or a semiconducting nanotube
of finite length the positions of energy levels around the energy
gap can be approximated with a good accuracy. The electron
spectrum of 0D carbon structures often include additional states
at energy $\varepsilon=0$, which are localized on zigzag edges and
do not contribute to the volume conductivity.
\end{abstract}


\maketitle


Carbon based materials of nano-size in the form of carbon
nanotubes (CNTs) are known for several years and attracted much
attention of researchers because of their unusual electronic
properties \cite{Iij,Dre,Sai}. Recently, progress in the
fabrication of other graphene-based lower dimensional structures
has been reported \cite{Nov1,Gei1}. This put forward such
nano-scaled quantum objects as one-dimensional (1D) carbon ribbons
(CR) \cite{Nak,Han,Ozy,Zhe,Raz} and zero-dimension (0D) carbon
dots \cite{Sco,Gei1,Ozy}.

Graphene - a two-dimensional (2D) carbon material - was first
isolated by micromechanical cleavage of graphite \cite{Nov2}. Its
planar hexagonal lattice is formed by $sp^2$ hybridized carbon
bonds. Although graphene is the building block of many carbon
allotropes, its electronic structure differs from other carbon
materials. At present, the electronic structure and transport
properties of 1D CNTs are well understood theoretically \cite{Cha}
and the focus is shifted towards CRs and 0D carbon objects
\cite{Mal}. In particular, it is known that in the framework of
the tight-binding model both CRs and CNTs (1D nano-materials), can
be either semiconducting with a size dependent gap or metallic
\cite{Nak,Han,Wak1}. It should be also noted that recent density
functional calculations within local density approximation
(DFT-LDA) predict that all armchair CRs are semiconducting, with
one group showing small energy gaps \cite{Son,Bar,Whi}.  We will
not discuss this issue here and limit our consideration by the
tight-binding (1D) and H\"{u}ckel model (0D). There, the rule of
metallicity for armchair CRs was formulated under assumption that
the ribbons are wide enough so that one can use solutions obtained
for graphene semiplane \cite{Wak1,Wak2}. Below we refine the
procedure and obtain the electronic solutions which are equally
applicable to CRs of small and large width. Furthermore, we show
that for the chosen class of 1D metallic carbon systems the
dispersion law of the electronic band crossing the Fermi level can
be obtained analytically [see Eq.~(\ref{C.9}) below]. Later we
generalize the law for 0D systems [Eq.~(\ref{C.12a}),
(\ref{C.12b}) below]. Throughout the letter we employ the
H\"{u}ckel model and limit ourselves to the class of zigzag CNTs
and CRs with the armchair profile. These two materials are closely
related with each other. Indeed, by rolling up an 1D armchair CR
one obtains a 1D zigzag CNT. (Terms armchair for the ribbon and
zigzag for the nanotube are confusing here since they apply to
different characteristics: armchair - to the edges of the ribbon,
while zigzag - to the circumference of the nanotube.) On the other
hand, all CRs of finite length can be considered as rectangular
graphene macromolecules (RGM) whose electronic properties are
important for designing various nano-materials.



%
\begin{figure}
\vspace{5mm} \resizebox{0.35\textwidth}{!} {

 \includegraphics{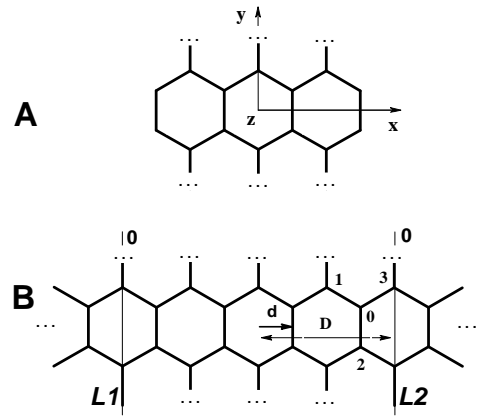}
}
\vspace{1mm} \caption{ Boundary conditions for a 1D carbon ribbon
(A) and the same ribbon placed in the graphene sheet (B). Carbon
sites on lines $L_1$ and $L_2$ have zero coefficients of
expansion. } \label{fig1}
\end{figure}
%
The energy spectrum of nanographite materials (1D CRs, 1D CNTs, 0D
RGM, 0D CNTs) can be obtained from the dispersion relation of
graphene \cite{Wal}:
\begin{eqnarray}
    \varepsilon_{+,-} = \varepsilon_F \pm \beta|\, e^{-iak_y \sqrt{3}/2} + 2 \cos ak_x/2 | ,
\label{C.1}
\end{eqnarray}
where $a = \sqrt{3} d_{CC}$, and $d_{CC}$ is the carbon-carbon
distance. Here the $2\pi$ factor is incorporated in $k$, and
$-\beta$ stands for the H\"{u}ckel transfer integral (or $-t$ in
the tight-binding model), so that $\beta>0$. At the $K-$point of
the Brillouin zone of graphene two bands intersect:
\begin{subequations}
\begin{eqnarray}
    & &a k_x^K = \pm \frac{4\pi}{3} , \label{C.2a} \\
    & &a k_y^K = 0 .  \label{C.2b}
\end{eqnarray}
\end{subequations}
We start with studying energy spectrum of 1D CRs. The problem is
how to reduce the problem to that for graphene. In the unit cell
of a zigzag carbon ribbon there are four carbon atoms with 2
nearest neighbors while in graphene each carbon atom has 3 nearest
neighbors, Fig.~\ref{fig1}. Therefore, if we consider the ribbon
as a part of the graphene 2D plane, Fig.~\ref{fig1}B, the
equations for the edge atoms of the ribbon are modified. For
example, for site $0$ in Fig.~\ref{fig1}B we have:
\begin{eqnarray}
    \frac{\varepsilon}{\beta} c_0 = -(c_1+c_2+c_3) ,
\label{C.3}
\end{eqnarray}
where $c_i$, $i=0,1,2,3$ are coefficients of the expansion of
graphene wave function $\psi_g$. However, if we put $c_3=0$, then
Eq.~(\ref{C.3}) will be identical to that for the ribbon. Thus, we
can select the solutions for the CR out of the graphene solutions
by requiting $c_j=0$ for all carbon sites $j$ on lines $L_1$ and
$L_2$ and by repeating the resulting pattern in the $x-$direction.
The lines with $\psi_r=0$ completely separates neighboring
nanoribbons from each other, since there is no interaction between
them.

We arrive at the following conditions of quantization:
\begin{subequations}
\begin{eqnarray}
    & &\cos( D k'_x )= 0 , \label{C.4a} \\
    & &\sin( D k'_x )= 0 . \label{C.4b}
\end{eqnarray}
\end{subequations}
Here $D=md$ ($m$ is an integer) and $d=\sqrt{3}d_{CC}/2$. (Both
distances are shown in Fig.~\ref{fig1}B.) The number of carbon
rings in the $x-$direction is ${\cal N} = m-1$. It is clear that
the CR will be metallic if line $(k'_x,k'_y)$ in $k-$space goes
through the $K-$point, Eqs.~(\ref{C.2a}), (\ref{C.2b}). Solving
Eqs.~(\ref{C.4a}), (\ref{C.4b}) leads to
\begin{eqnarray}
    ak'_x = 2\pi \frac{n}{m} ,
\label{C.6}
\end{eqnarray}
where $n$ is an integer number. Condition (\ref{C.6}) coincides
with Eq.~(\ref{C.2a}), (\ref{C.2b}) only if
\begin{subequations}
\begin{eqnarray}
    & &m = 3m' , \label{C.7a} \\
    & &n = 2m' . \label{C.7b}
\end{eqnarray}
\end{subequations}
This immediately gives
\begin{eqnarray}
    {\cal N} = 3m'-1 .
\label{C.8}
\end{eqnarray}
Eq.~(\ref{C.8}) determines which CR is metallic. We get: ${\cal
N}$=2, 5, 8,... . Substituting $(k'_x=k_x^K,k'_y=k)$ in
Eq.~(\ref{C.1}) leads to the dispersion relations for two bands
which cross at the Fermi energy:
\begin{eqnarray}
     \varepsilon_{+,-} = \varepsilon_F \pm 2 \beta \sin \frac{X k}{4},
\label{C.9}
\end{eqnarray}
where $X=3d_{CC}$ ($X$ is the modulus of the basis vector in the
$y-$direction). The corresponding density of states (DOS) {\it per
unit cell} (for both spins projections) in the neighborhood of
$\varepsilon_F=0$ is given by
\begin{eqnarray}
    \rho(\varepsilon) = \frac{2}{\pi}\, \frac{1}{\sqrt{ 1-(\varepsilon/2 \beta)^2}}\,
    \frac{1}{\beta} .
\label{C.9'}
\end{eqnarray}
Eq.~(\ref{C.9'}) coincides with that for metallic CNTs \cite{Cha}.

\begin{figure}
\vspace{5mm} \resizebox{0.25\textwidth}{!} {

 \includegraphics{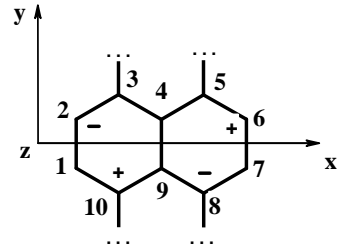}
}
\vspace{1mm} \caption{ Unit cell of 1D metallic carbon ribbon.
Plus signs (sites 6, 10) and minus signs (2, 8) refer to
eigenvector $V_1$ at $\varepsilon_F=0$.} \label{fig2}
\end{figure}

It is instructive to study the two electronic states at the Fermi
energy, when $k_F=0$. We start by considering the simplest
possible case: ${\cal N}=2$, Fig.~\ref{fig2}. Then the following
explicit eigenvectors can be found:
\begin{subequations}
\begin{eqnarray}
    V_1 = \frac{1}{2}\, \left\{ 0, -1, 0, 0, 0, 1, 0, -1, 0, 1 \right\} , \label{1.4a} \\
    V_2 = \frac{1}{2}\, \left\{ -1, 0, 1, 0, -1, 0, 1, 0, 0, 0 \right\}. \label{1.4b}
\end{eqnarray}
\end{subequations}
$V_1$ is visualized in Fig.~\ref{fig1} by putting plus and minus
signs at corresponding carbon sites. $V_2$ is obtained from $V_1$
through the mirror reflection at the $xz$ plane, which follows
from the symmetry of the electronic system.
%
%
%
\begin{figure}
\vspace{5mm} \resizebox{0.3\textwidth}{!} {

 \includegraphics{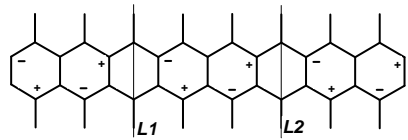}
}
\vspace{1mm} \caption{ Unit cell of 1D metallic carbon ribbon
containing 8 carbon rings. Plus and minus signs schematically
indicate one of two eigenvectors at $\varepsilon_F=0$.}
\label{fig3}
\end{figure}
%
In this way one can construct two eigenvalues for any metallic CR.
As an example in Fig.~\ref{fig3} we schematically draw one
eigenvector for a CR with the 8 ring unit cell. Notice, that all
three blocks have the structure of $V_1$, Fig.~\ref{fig2}. The
blocks are connected via chains of carbon sites - lines $L_1$ and
$L_2$ - where $c_j = 0$. As before, the second eigenvector is
obtained from the first through the $xz-$mirror reflection. Thus,
we can build two eigenvectors only if number of blocks is ${\cal
N}=2+3n$, which also gives the rule (\ref{C.8}).

From Eq.~(\ref{C.1}) one can derive the expression for energy gap
$E_g$ if ${\cal N} \neq 3m'-1$. There are two cases:(1) ${\cal N}
= 3m'$ and $m=3m'+1$, and (2) ${\cal N} = 3m'-2$, $m=3m'-1$. In
both cases the energy gap is given by
\begin{subequations}
\begin{eqnarray}
     E_g = 2 \beta\, |1 - \cos \frac{\delta \phi}{2} - \sqrt{3} \sin \frac{\delta
     \phi}{2}| ,
     \label{C.10a}
\end{eqnarray}
The value of $\delta \phi$ depends on which branch -
Eq.~(\ref{C.4a}) or Eq.~(\ref{C.4b}) - is considered. [It is a
measure of deviation from the $K-$point, Eq.~(\ref{C.2a},b) in the
zone-folding model.] For {\it cos}-like branches,
Eq.~(\ref{C.4a}), we get $n=2m'$, Eq.~(\ref{C.7b}), and
\begin{eqnarray}
      \delta \phi_{cos} = \mp \frac{\pi}{3({\cal N} + 1)} .
     \label{C.10b}
\end{eqnarray}
Here the first choice of ${\cal N}$ corresponds to minus sign,
while the second choice of ${\cal N}$ to plus sign. For {\it
sin}-like branches, Eq.~(\ref{C.4a}), we get
\begin{eqnarray}
    & &\delta \phi_{sin} = 4 \phi_{cos}, \quad n=2m' \label{C.10c} \\
    & &\delta \phi_{sin} = 2 \phi_{cos}, \quad n=2m'\pm1 \label{C.10d}
\end{eqnarray}
\end{subequations}
By comparing Eq.~(\ref{C.10b}) with Eq.~(\ref{C.10c}) and
Eq.~(\ref{C.10d}) we conclude that the gap is due to {\it
cos-}like branches, Eq.~(\ref{C.10b}). If $\delta \phi_{cos} \ll
1$, which is often the case,
\begin{eqnarray}
     E_g =  \beta\, \frac{4\pi}{\sqrt{3}} \frac{1}{{\cal N} + 1} .
     \label{C.11}
\end{eqnarray}
Thus, $E_g \sim 1/{\cal N}$ as it was the case for CNTs
\cite{Cha}.

The rule for the metallicity of armchair CRs, Eq.~(\ref{C.8}), is
very different from the the rule for the corresponding zigzag CNTs
characterized by the pair of indices $[{\cal N},0]$:
\begin{eqnarray}
    {\cal N} = 3m' .
\label{C.12}
\end{eqnarray}
It is clear that the rules do not overlap meaning that if one
rolls up a nanotube from a metallic ribbon, then the resultant
nanotube will not be metallic and vice versa. This conclusion
deserves a more detailed explanation. First, we notice that the
metallicity rule for CNTs, Eq.~(\ref{C.12}), is obtained from the
cyclic condition in the $x-$direction which allows for sin- and
cos- like dependencies, while in case of ribbon only sin- like
functions are allowed. [One can prove that Eq.~(\ref{C.4a}) does
not include the $K-$point, Eq.~(\ref{C.2a}).] The sin-dependence
in the $x-$direction implies opposite signs for coefficients $c_j$
belonging to edge sites $j$ on opposite sides of CR. By rolling up
a nanotube the carbon sites on opposite edges of ribbon should
coincide, which in turn destroys the odd solutions. The even
solutions satisfying the cyclic condition survive the rolling up
procedure but none of them has energy at $\varepsilon_F=0$.

We now turn to 0D objects - carbon macromolecules and nanotubes of
finite length. Unlike 1D electron systems characterized by
electron energy band structure, they have discrete energy spectra.
However, these spectra are closely related with electron bands
which we have already considered. In particular, energy levels of
RGMs and CNTs near the Fermi energy are described by the following
expression:
\begin{subequations}
\begin{eqnarray}
     \varepsilon_{+,-}(n) = \pm 2 \beta \sin \frac{X k(n)}{4},
\label{C.12a}
\end{eqnarray}
where
\begin{eqnarray}
    k(n) = \delta k\, \left(n + \frac{1}{2} \right) ,
\label{C.12b}
\end{eqnarray}
\end{subequations}
$n=0,1,2$,... and $\delta k = \pi/L$, and $L=L_0 + 7d_{CC}/4$.
Here $L_0$ is the length of nanotube or ribbon (maximal distance
between carbon atoms in the $y-$direction, which has the armchair
profile). We want to stress that Eqs.~(\ref{C.12a}), (\ref{C.12b})
are {\it exact}, see Appendix. The electron spectrum given by
(\ref{C.12a}) is independent of width which is consistent with the
situation observed for 1D objects. It is also worth noting that
the spectrum of 0D carbon objects consists of many discrete levels
and instead of Fermi level we should speak of highest occupied
(HOMO) and lowest unoccupied (LUMO) molecular orbitals. This
implies that formally we can not speak of metallicity and
semiconductivity. However, in Tables~\ref{tab1}-\ref{tab3} we
retain these terms in a loose sense, because from one side, it
shows relations with corresponding 1D objects and from the other
side, the existence or nonexistence of a large energy gap at the
HOMO-LUMO region remains one of the important characteristics of
these systems.
\begin{table}

\caption{ Length dependence of the energy spectrum of metallic
RGMs (${\cal N}=2,5,...$) and CNTs (${\cal N}=3,6,...$).
\label{tab1} }

\begin{tabular}{c c c c}

\hline
         & \multicolumn{3}{c}{ length, in $d_{CC}$ } \\
HOMO$-i$ & 299    &  149     & 29     \\
\hline
1 & -0.00783 & -0.01563 & -0.07661 \\
2 & -0.02350 & -0.04689 & -0.22937 \\
3 & -0.03917 & -0.07813 & -0.38078 \\
4 & -0.05483 & -0.10935 & -0.52996 \\
5 & -0.07049 & -0.14055 & -0.67603 \\
6 & -0.08615 & -0.17172 & -0.81814 \\
7 & -0.10180 & -0.20284 & -0.95544 \\

\hline

\end{tabular}
\end{table}
\begin{table}

\caption{ Length dependence of the energy spectrum of
semiconducting RGM, ${\cal N}=10$. ($L_0=3.5d_{CC}$,
$\delta=0.9$). ``Exact" refers to the strait-forward H\"{u}ckel
calculations, ``approx" to the values of Eq.~(\ref{C.14a},b).
\label{tab2} }

\begin{tabular}{c c c c c}

\hline
         & \multicolumn{2}{c}{299 $d_{CC}$}    &  \multicolumn{2}{c}{214 $d_{CC}$}      \\
HOMO$-i$ & exact & approx. & exact & approx. \\
\hline
1 & -0.08271 &  -0.08258 &   -0.08371 & -0.08356 \\
2 & -0.08689 &  -0.08691 &   -0.09069 & -0.09100 \\
3 & -0.09349 &  -0.09388 &   -0.10143 & -0.10250 \\
4 & -0.10210 &  -0.10294 &   -0.11501 & -0.11687 \\
5 & -0.11230 &  -0.11360 &   -0.13064 & -0.13318 \\
6 & -0.12374 &  -0.12544 &   -0.14771 & -0.15079 \\
7 & -0.13611 &  -0.13816 &   -0.16580 & -0.16928 \\

\hline

\end{tabular}
\end{table}
\begin{table}

\caption{ Length dependence of the energy spectrum of
semiconducting CNT $[10,0]$ (${\cal N}=10$, $L_0=3.75d_{CC}$,
$\delta=0.9$). ``Exact" refers to the strait-forward H\"{u}ckel
calculations, ``approx" to the values of Eq.~(\ref{C.14a},b).
\label{tab3} }

\begin{tabular}{c c c c c}

\hline
         & \multicolumn{2}{c}{299 $d_{CC}$}    &  \multicolumn{2}{c}{214 $d_{CC}$}      \\
HOMO$-i$ & exact & approx. & exact & approx. \\
\hline
1 & -0.17634 &  -0.17623 & -0.17691 &  -0.17673 \\
2 & -0.17864 &  -0.17847 & -0.18088 &  -0.18067 \\
3 & -0.18241 &  -0.18226 & -0.18733 &  -0.18723 \\
4 & -0.18756 &  -0.18750 & -0.19602 &  -0.19615 \\
5 & -0.19400 &  -0.19406 & -0.20671 &  -0.20712 \\
6 & -0.20161 &  -0.20182 & -0.21910 &  -0.21982 \\
7 & -0.21027 &  -0.21065 & -0.23296 &  -0.23398 \\

\hline

\end{tabular}
\end{table}

It is clear that Eqs.~(\ref{C.12a}) and (\ref{C.12b}) can be
considered as a discretization of (\ref{C.9}). Following this
route we can derive an approximate expression for nonmetallic RGMs
and CNTs. First, we recall that for 1D systems there are various
bands, which we have discussed already while calculating $E_g$. In
general their energy is
\begin{eqnarray}
     \frac{\varepsilon_{+,-}}{\beta} = \pm
     | e^{-i\frac{X}{2} k} - \cos \frac{\delta \phi}{2} - \sqrt{3} \sin \frac{\delta
     \phi}{2} |,
      \label{C.12'}
\end{eqnarray}
where $\delta \phi = \delta \phi_{sin}$ or $\delta \phi_{cos}$.
The highest occupied band for RGMs is given by $\delta
\phi_{cos}$, Eq.~(\ref{C.10b}). Starting with Eq.~(\ref{C.12'})
and taking inspiration from the relation between (\ref{C.9}) and
(\ref{C.12a}) we obtain the discretized version for
$\varepsilon_{+,-}(n)$ of semiconducting RGMs or CNTs of finite
length:
\begin{subequations}
\begin{eqnarray}
     & &\frac{\varepsilon_{+,-}(n)}{\beta} = \pm
     | e^{-i\frac{X}{2} k'(n)} - \cos \frac{\delta \phi}{2} - \sqrt{3} \sin \frac{\delta
     \phi}{2} |, \quad \quad
      \label{C.14a} \\
    & &k'(n) = \delta k\, \left(n + \delta \right)  .
\label{C.14b}
\end{eqnarray}
\end{subequations}
Here $n=0,1,2$,... is integer, $\delta k =\pi/L$, and
$L=L_0+\delta L$. In fact, in Eq.~(\ref{C.14b}) $\delta L \sim
d_{CC}$ and $\delta \sim 1$ are phenomenological parameters which
should be found by fitting a data set of $\varepsilon(n)$. The set
can be taken from a H\"{u}ckel calculation of RGMs and CNTs or
even from a more advanced calculation (like density functional,
for example). In Tables~\ref{tab2} and \ref{tab3} we compare
energy spectra given by Eq.~(\ref{C.14a},b) [approx.] with
straight-forward H\"{u}ckel calculations [exact]. $\delta \phi$ in
Eq. (\ref{C.14a}) is $\phi_{cos}$, Eq.\ (\ref{C.10b}), for
graphene molecules, and
\begin{eqnarray}
    \delta \phi = \pm \frac{2 \pi}{3 {\cal N}}
\label{C.17}
\end{eqnarray}
in case of semiconductiong nanotubes with width ${\cal N}= 3m' \pm
1$. The accuracy of Eqs.~(\ref{C.14a}), (\ref{C.14b}) is of the
order of $10^{-3} \beta$ for 10 top occupied molecular orbitals
(or HOMO$-i$, $i=1-10$), Tables~\ref{tab2}, \ref{tab3}.

\begin{figure}
\vspace{-0mm} \resizebox{0.49\textwidth}{!} {

 \includegraphics{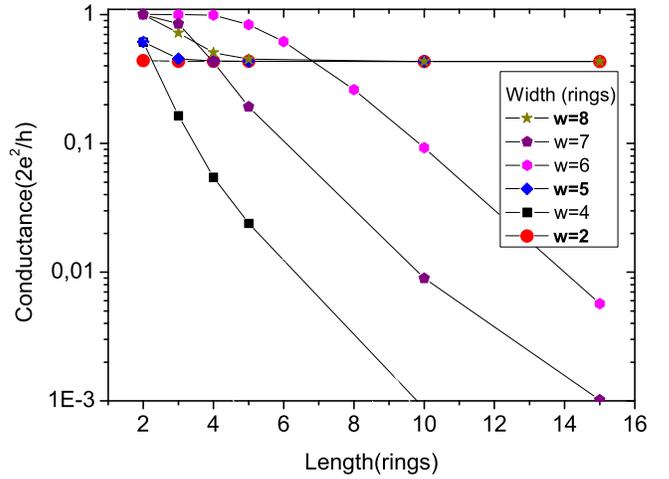}
}
\vspace{-6mm} \caption{ Conductance versus RGM length at
$\varepsilon_F=0$. } \label{fig4}
\end{figure}

Finally, we remark on the doubly degenerate HOMO level at
$\varepsilon \approx 0$ \cite{Nak,Wak1,Wak2}. The nature of the
states which are always present in the electronic spectrum of
carbon ribbons and nanotubes of finite length, is an object of
intense research \cite{Nak,Wak1,Wak2,Mun,Sas,Son2,Hod}. These
states are edge ones, because as follows from Eqs.~(\ref{C.12a}),
(\ref{C.12b}) $\varepsilon_{HOMO-1}=-2\beta \sin (X\delta k/8)
\neq 0$. The coefficients of the wave function expansion, $c_j$,
in this case quickly fall to zero as we move away from the edges,
and the states do not contribute to the conductivity in the
$y-$direction, Fig.~\ref{fig1}. To demonstrate it we have
calculated the conductance of RGMs using the Landauer formula
\cite{Imry}. For a qualitative treatment, we define the self
energy within the broadband approximation~\cite{Muj}, considering
it as an energy-independent imaginary constant, $\Sigma=i\Delta$
($\Delta=1$~eV). In Fig.~\ref{fig4} we plot conductance as a
function of length at $\varepsilon_F=0$ for RGMs of various width.
Our results show that metallic RGMs ($n=2,5,8$) have a weak
dependence on length. Their conductances coincide starting with a
rather short length of 2-6 rings. For other non-metallic RGMs the
calculations demonstrate an exponential decrease of conductance
with length. It is also worth noting that the conductance of
metallic RGMs is always the same and equal to the conductance of
the metallic RGM with the minimal width of 2 rings. (A detailed
qualitative and quantitative analysis will be given elsewhere.)
%
%

In conclusion, we have studied the electronic spectrum of the 1D
armchair CRs and zigzag CNTs and their 0D counterparts of finite
length in the H\"{u}ckel model. We have found the solutions by
reducing the problem to that for graphene with appropriate
selection rules imposed by boundary conditions. In the vicinity of
the HOMO-LUMO energy region ($\varepsilon \sim 0$) we have found
the exact expression for energy spectrum of metallic nanosystems,
Eq.~(\ref{C.12a},b) and approximate energy spectrum in case of
semiconducting materials, Eq.~(\ref{C.14a},b). Finally, we have
calculated the conductance of some RGMs and investigated the role
of edge states.

The authors would like to thank D. S. Kosov for helpful
discussions.

\renewcommand{\theequation}{A-\arabic{equation}}
\setcounter{equation}{0}
\section*{APPENDIX}

\begin{figure}
\vspace{-0mm} \resizebox{0.4\textwidth}{!} {

 \includegraphics{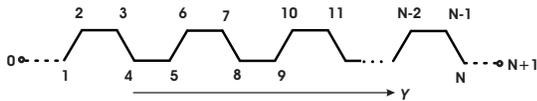}
}
\vspace{0mm} \caption{ Effective 1D chain of $N$ carbon atoms
[$i=1,2,...,N$] within the $l$-ribbon. Also shown are two
auxiliary carbon atoms [$i=0$ and $i=(N+1)$], see text for
details. } \label{fig5}
\end{figure}

Here we discuss the derivation of Eq.~(\ref{C.12a},b). For a 1D
armchair CR or 1D zigzag CNT each of the discrete values of $k'_x$
determines an electronic band. The metallic band is given by
Eq.~(\ref{C.2a}) and obtained for the (\ref{C.4b}) condition of
quantization. This condition implies the $\sin[(2\pi/3)(x/d)]$
modulation in the $x-$direction with zero coefficients at $x/d=3l$
where $l=0,1,2, ...$. (Distances and axes are shown in
Fig.~\ref{fig1}.) Each index $l$ defines a carbon ribbon in the
$y-$direction with nonzero coefficients at $x_{1}/d=1+3l$ and
$x_{2}/d=2+3l$. The lines with zero coefficients imply that each
ribbon can be considered as independent. In fact, they are
equivalent due to the modulation condition, Eq.~(\ref{C.4b}). We
can use this property and work only with one ribbon shown in
Fig.~\ref{fig5} and later reconstruct the solution for the whole
system.
%
%

Now we consider one ribbon in the $y-$direction which is
equivalent to a one dimensional (1D) chain of carbon atoms
($i=1,2,...,N$). Considering a solution with the coefficients
$c_j$ of the wave function expansion, we obtain for them the
following relations
\begin{eqnarray}
    \frac{\epsilon}{\beta} c_j = -(c_{j-1}+c_{j+1}) ,
\label{a.1}
\end{eqnarray}
where $j=2,3,...,(N-1)$. The problem arises due to the two
boundary atoms: $j=1$ and $j=N$. To solve this task we will use
the trick which we have already applied to 1D ribbon with
Eq.~(\ref{C.3}). That is, we introduce two auxiliary carbon atoms
$j_{b1}=0$ and $j_{2b}=(N+1)$ and consider an infinite 1D carbon
chain beyond them. Notice, that for the following the real shape
of the 1D chain is immaterial, it can be equally thought of as a
1D linear chain of carbon atoms. For an infinite chain the general
solution is $c_j \sim exp(-\phi'\, j)$, where $\phi'=d_{CC}\,
k'_y$ and $k'_y$ is an effective wave number. Then the
coefficients at $j_b$ should be zero, i.e. $c_{0}=c_{N+1}=0$. This
gives a condition of quantization,
\begin{eqnarray}
    \sin(L_y\, k'_y) = 0,
\label{a.2}
\end{eqnarray}
where $L_y=(N+1)\, d_{cc}$. From (\ref{a.2}) we get
\begin{eqnarray}
    k'_y = \frac{\pi}{L_y}\, r = \frac{\pi}{(N+1)\, d_{CC}}\, r .
\label{a.3}
\end{eqnarray}
Here positive integer $r=1,2,...,N$. The energy of the 1D chain is
obtained from Eq.~(\ref{a.1}):
\begin{eqnarray}
    \epsilon/\beta = -2\cos(\phi')=-2\sin \left(\frac{\pi}{2} -
    \phi'\right).
\label{a.4}
\end{eqnarray}
The latter relation can be written as
\begin{eqnarray}
    \epsilon/\beta = -2\sin \left(\frac{\pi}{N+1}(n + \frac{1}{2})\right),
\label{a.5}
\end{eqnarray}
where
\begin{eqnarray}
    2n = N - 2r.
\label{a.6}
\end{eqnarray}
Notice that $n=N/2-1$, $(N/2-2)$, ..., $-N/2$. ($N$ is even,
because $N=4N_{hex}$, where $N_{hex}$ is the number of hexagons in
the $y-$direction.) First $N/2$ electron states are occupied by
$N$ $\pi$-electrons. Taking into account that the length of CR or
CNT in the $y-$direction,
\begin{eqnarray}
    L_0 = \left( \frac{3}{4}N -1 \right)\, d_{CC} ,
\label{a.7}
\end{eqnarray}
Eq.~(\ref{a.5}) can be rewritten as
\begin{eqnarray}
     \epsilon/\beta = \mp 2\sin \left(\frac{3d_{CC}}{4} \frac{\pi}{(L_0+7d_{CC}/4)}(n' + \frac{1}{2})\right),
\label{a.8}
\end{eqnarray}
where the solution with the minus sign refers to the occupied
states at the Fermi level: HOMO ($n'=0$), and the HOMO$-n'$ levels
($n'=1,2,...,N/2$.) The solution with the plus sign refers to the
unoccupied states: LUMO ($n'=0$), and the LUMO$+n'$ levels
($n'=1,2,...,N/2$.). Eq.~(\ref{a.8}) is equivalent to
Eq.~(\ref{C.12a},b). Finally, we would like to notice that the
equation implies a certain size relation of RGM, $L_y=L_0>L_x$.
The latter condition is needed to assure that a set of levels
around the Fermi level is associated with a 1D metal band and
separated from other $k'_x-$levels (other 1D bands). If it is not
so ($L_y=L_0<L_x$) then the levels described by (\ref{a.8}) do
exist but they are not grouped together. They are mixed up with
other $k'_x-$levels.



\end{document}